\documentclass[%
preprint,
superscriptaddress,
 amsmath,amssymb,
 aps,
 prb,
]{revtex4-2}

\usepackage[colorlinks, linkcolor=blue, anchorcolor=blue, citecolor=blue, urlcolor=blue]{hyperref} 
\usepackage{graphicx}
\usepackage{dcolumn}
\usepackage{bm}
\usepackage{float}
\usepackage{multirow}
\usepackage{setspace}

\begin{document}

\title{Electrical switching of ferro-rotational order in nano-thick 1\textit{T}-TaS$_2$ crystals}

\author{Gan~Liu}
\thanks{These authors contributed equally to this work}
\author{Tianyu~Qiu}
\thanks{These authors contributed equally to this work}
\author{Kuanyu~He}
\thanks{These authors contributed equally to this work}
\affiliation{National Laboratory of Solid State Microstructures and Department of Physics, Nanjing University, Nanjing 210093, China}

\author{Yizhou~Liu}
\affiliation{Department of Condensed Matter Physics, Weizmann Institute of Science, Rehovot 7610001, Israel}

\author{Dongjing~Lin}
\affiliation{National Laboratory of Solid State Microstructures and Department of Physics, Nanjing University, Nanjing 210093, China}

\author{Zhen~Ma}
\affiliation{National Laboratory of Solid State Microstructures and Department of Physics, Nanjing University, Nanjing 210093, China}
\affiliation{Institute for Advanced Materials, Hubei Normal University, Huangshi 435002, China}
\affiliation{State Key Laboratory of Surface Physics and Department of Physics, Fudan University, Shanghai 200433, China}

\author{Zhentao~Huang}
\affiliation{National Laboratory of Solid State Microstructures and Department of Physics, Nanjing University, Nanjing 210093, China}

\author{Wenna~Tang}
\author{Jie~Xu}
\affiliation{National Laboratory of Solid State Microstructures and Department of Physics, Nanjing University, Nanjing 210093, China}

\author{Kenji Watanabe}
\affiliation{Research Center for Functional Materials, 
National Institute for Materials Science, 1-1 Namiki, Tsukuba 305-0044, Japan}

\author{Takashi~Taniguchi}
\affiliation{International Center for Materials Nanoarchitectonics, National Institute for Materials Science,  1-1 Namiki, Tsukuba 305-0044, Japan}

\author{Libo~Gao}
\author{Jinsheng~Wen}
\author{Jun-Ming Liu}
\affiliation{National Laboratory of Solid State Microstructures and Department of Physics, Nanjing University, Nanjing 210093, China}
\affiliation{Collaborative Innovation Center of Advanced Microstructures, Nanjing University, Nanjing 210093, China}

\author{Binghai~Yan}
\email{binghai.yan@weizmann.ac.il}
\affiliation{Department of Condensed Matter Physics, Weizmann Institute of Science, Rehovot 7610001, Israel}

\author{Xiaoxiang~Xi}
\email{xxi@nju.edu.cn}
\affiliation{National Laboratory of Solid State Microstructures and Department of Physics, Nanjing University, Nanjing 210093, China}
\affiliation{Collaborative Innovation Center of Advanced Microstructures, Nanjing University, Nanjing 210093, China}

\maketitle

\noindent
\textbf{Abstract}

\noindent
\textbf{Hysteretic switching of domain states is a salient character of all ferroic materials and the foundation for their multifunctional applications. Ferro-rotational order is emerging as a new type of ferroic order featuring structural rotations, but its controlled switching remains elusive due to its invariance under both time reversal and spatial inversion. Here, we demonstrate electrical switching of ferro-rotational domain states in nanometer-thick 1\textit{T}-TaS$_2$ crystals in its charge-density-wave phases. Cooling from the high-symmetry phase to the ferro-rotational phase under an external electric field induces domain state switching and domain wall formation, realized in a simple two-terminal configuration using a volt-scale voltage. Although the electric field does not couple with the order due to symmetry mismatch, it drives domain wall propagation to give rise to reversible, durable, and nonvolatile isothermal state switching at room temperature. These results pave the path for manipulation of the ferro-rotational order and its nanoelectronic applications.}

\vspace{10mm}
\noindent
\textbf{Main}

Ferroic orders arise from symmetry-breaking phase transitions, finding applications in wide-ranging advanced technologies~\cite{Wadhawan2000}. Symmetry therefore provides a powerful guide to the identification of conjugate fields that couple with and even switch the orders --- a prerequisite for utilizing the associated multi-stable domain states~\cite{Cheong2018}. This principle is well applicable to three out of the four types of ferroics with a vector order parameter~\cite{Cheong2018}: ferromagnets (ferroelectrics) feature time-reversal (spatial-inversion) symmetry breaking spontaneous magnetization (electric polarization) that is switchable by a magnetic (electric) field, whereas the ferro-toroidal order breaks both symmetries and can be switched using composite magnetic and electric fields~\cite{Aken2007,Spaldin2008,Zimmermann2014}. The remaining type --- the ferro-rotational (also known as ferroaxial) order --- stands out, as it is both time-reversal and spatial-inversion invariant, hence insensitive to electromagnetic fields~\cite{Cheong2018,Hlinka2016}. The lack of external fields that hysteretically switch the ferro-rotational order casts doubts on its ferroic nature~\cite{Fiebig2020} and limits its potential applications.

\begin{figure*}[t]
\centering
\includegraphics[width=1\linewidth]{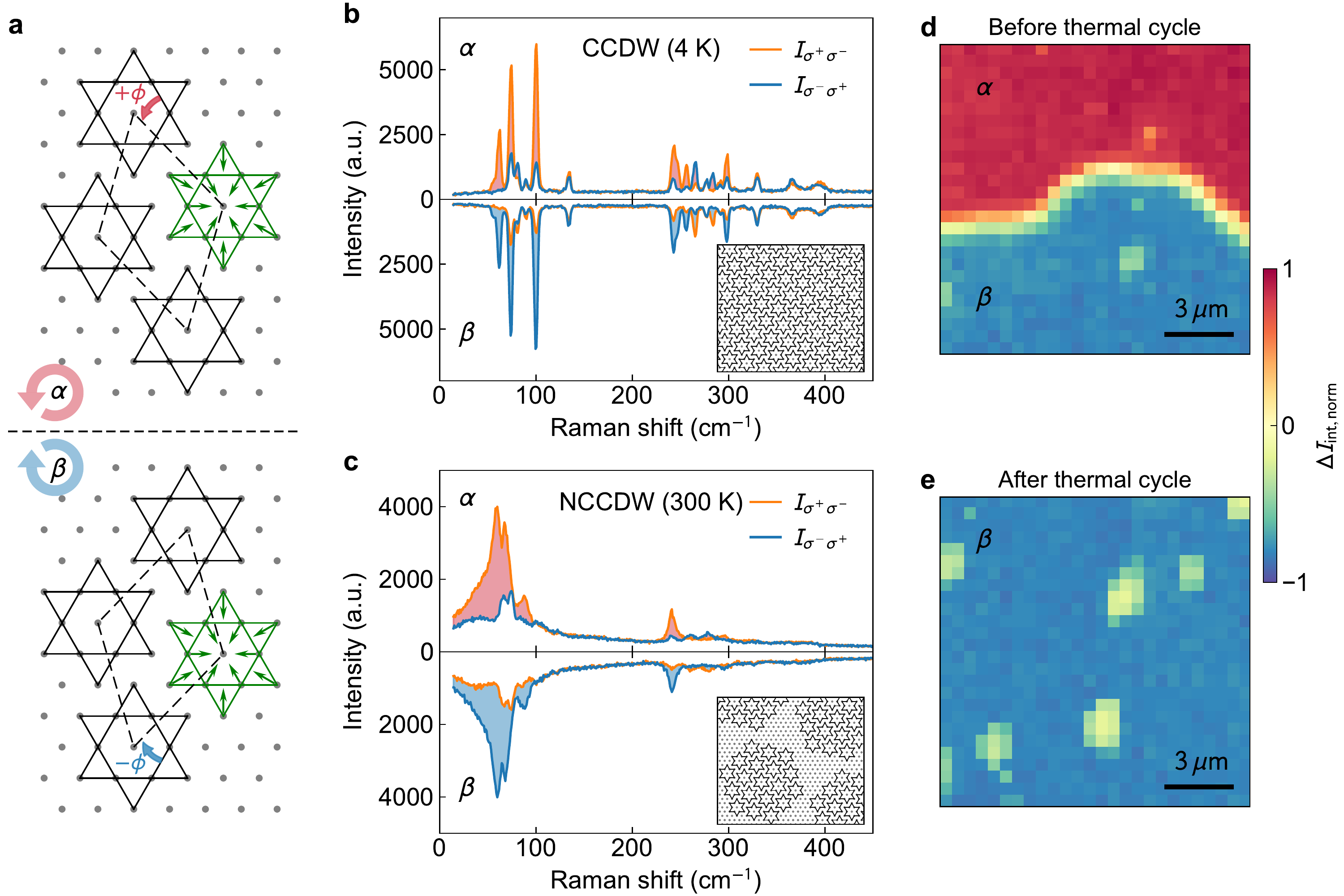}
\caption{\textbf{Ferro-rotational order in 1\textit{T}-TaS$_2$ and its helicity-resolved Raman response.} \textbf{a}, Green: schematic of the star-of-David (SD) lattice distortion, with 12 peripheral Ta atoms displaced towards a central Ta atom to form a cluster. Two ways of tiling the $ab$-plane with this structural motif form planar-chiral configurations, $\alpha$ and $\beta$, with the superlattice unit cell (dashed rhombus) rotated 13.9$^{\circ}$ counterclockwise and clockwise, respectively, relative to the atomic lattice. \textbf{b}, \textbf{c}, Raman spectra for two bulk samples with opposite chiral states in the CCDW and NCCDW phases (illustrated in the insets). The contrast between the $\sigma^+\sigma^-$ and $\sigma^-\sigma^+$ channels (filled colors) distinguishes the planar chirality. \textbf{d}, \textbf{e}, Raman mapping of the normalized integrated contrast $\Delta I_{\mathrm{int, norm}}$ for a bulk sample at 300~K, before and after warming to the ICCDW phase. }
\label{Fig1}
\end{figure*}

Ferro-rotational order is attracting growing interest as a new type of ferroic order induced by rotational structural transitions~\cite{Gopalan2011,Johnson2011,Johnson2012,Hlinka2016,Cheong2018,Jin2020,Luo2021,Hayashida2020,Hayashida2021,Cheong2021}. It possesses an axial-vector order parameter that distinguishes its domain states\cite{Hlinka2016}. Coupled with magnetism, it can give rise to multiferroic properties~\cite{Johnson2011,Johnson2012}. Similar order has been observed in nanoscale ferroelectrics with electric toroidal moments~\cite{Naumov2004,Damodaran2017}. Recent advances have enabled detection of the ferro-rotational order and domain imaging in a variety of materials, using second harmonic generation and linear electro-gyration~\cite{Fichera2020,Jin2020,Luo2021,Hayashida2020,Hayashida2021}. Here we focus on a pure ferro-rotational material candidate, 1\textit{T}-TaS$_2$, which does not couple with other types of ferroic orders. The van der Waals layered dichalcogenide 1\textit{T}-TaS$_2$ is a prototypical charge-density-wave (CDW) system exhibiting rich phase diagrams and high tunability~\cite{Sipos2008,Stojchevska2014,Yoshida2014,Yu2015,Yoshida2015,Cho2016,Ma2016,Vaskivskyi2016,Qiao2017}. In the CDW phases, the distorted lattice features star-of-David (SD) clusters formed by 13 Ta atoms (Fig.~\ref{Fig1}a), essential for the insulating state~\cite{Fazekas1980}. Tiling of SD clusters leads to degenerate configurations, $\alpha$ and $\beta$, with the superlattice rotated counterclockwise and clockwise, respectively, relative to the atomic lattice~\cite{Wilson1975,Thomson1994}. The structure is planar chiral~\cite{Zong2018,Fichera2020}, breaking the vertical mirror symmetry that relates the two configurations. This peculiar feature, largely overlooked in the past decades, has come to the fore in the pursuit of chiral CDWs~\cite{Ishioka2010,Xu2020,Jiang2021,Sung2022,Yang2022} and is susceptible to optical~\cite{Zong2018} or electrical~\cite{Song2022} pulses. Recently, the planar-chiral phase was recognized to be ferro-rotational in nature~\cite{Luo2021}, hosting two domain states corresponding to $\alpha$ and $\beta$. However, hysteretic switching between the domain states, which is a generic hallmark of all ferroics~\cite{Wadhawan2000}, remains elusive.

In this Article, using helicity-resolved Raman spectroscopy and imaging, we demonstrate unexpected remarkable macroscopic switching of the structural rotation in nanometer-thick 1\textit{T}-TaS$_2$ via a linear electric field. Although the distorted lattice induced by the CDW transition preserves global inversion symmetry and precludes coupling with an external electric field, the local electric dipoles at the domain wall induced by the CDW distortion can respond to the field, which in turn alters the local distortion. This leads to the expansion of one type of domain at the expense of the other, and hence the state switching. Our results open a new perspective to domain state switching, emphasizing domain wall propagation as the key driving force, which can be triggered by external fields that do not necessarily couple with the order itself. The switchable ferro-rotational order hosts applications in non-volatile memories and optoelectronics. Since 1\textit{T}-TaS$_2$ is known to harbor electron correlation and superconductivity, the switchable planar-chiral lattice offers a new approach to engineer novel states, especially at the domain wall. It also provides a platform to explore tunable nonreciprocal effects~\cite{Cheong2018,Cheong2021}.

\vspace{2mm}
\noindent
\textbf{Detection of ferro-rotational order}

Upon cooling, bulk 1\textit{T}-TaS$_2$ undergoes successive CDW transitions, entering an incommensurate phase (ICCDW) at around 550~K, a nearly-commensurate phase (NCCDW) at 350~K, and a commensurate phase (CCDW) at 180~K~\cite{Wilson1975}. Close packing of SD clusters corresponds to the CCDW phase (Fig.~1b inset), leading to a long-range ferro-rotational order. The NCCDW phase consists of hexagonal patches of such distortions with coherent planar chirality, separated by diffuse discommensurations (Fig.~1c inset)~\cite{Wu1989,Thomson1994,Spijkerman1997,Park2019}, whereas in the ICCDW phase SD clusters are absent. Therefore ferro-rotational order is retained in the former but vanishes in the latter.

Planar chiral crystals are endowed with Raman tensors with off-diagonal components, giving rise to an interesting helicity-resolved Raman response that enables efficient discrimination of the two configurations~\cite{Yang2022}. Since ferro-rotational materials must also be planar chiral (Supplementary Note 1), this establishes helicity-resolved Raman scattering as a new probe of ferro-rotational order. Figures~1b and~1c compare the Raman spectra $I_{\sigma^+\sigma^-}$ and $I_{\sigma^-\sigma^+}$ for two samples with opposite planar chirality, where $\sigma^i\sigma^s$ $(i,s = +,-)$ denotes helicities for the incident and scattered photons. Each sample shows nonzero differential spectra $\Delta I \equiv I_{\sigma^+\sigma^-}- I_{\sigma^-\sigma^+}$ in the CCDW and NCCDW phases for the $E_g$ phonons~\cite{Yang2022}, whereas $\Delta I =0$ in the achiral ICCDW phase (Supplementary Fig.~1). Moreover, the two states are related by $(\Delta I)_{\alpha}=- (\Delta I)_{\beta}$. While the correspondence between the domain states and Raman response is yet to be determined, we assume that the $\alpha$ ($\beta$) state corresponds to positive (negative) $\Delta I_{\mathrm{int}}$, the differential spectrum integrated over 15--150~cm$^{-1}$.

We tested dozens of bulk flakes with dimensions up to tens of microns and found most of them monodomain. This is distinct from other types of ferroics, in which multiple domains coexist to compensate surface fields and minimize surface energy~\cite{Catalan2012}. Ideally, cooling from the ICCDW phase to the NCCDW phase should realize the two states with equal probability. Instead, we found predetermined domain state for a given homogeneous sample probably due to extrinsic factors~\cite{Zong2018}, which remained unaltered by thermal cycling. On one rare sample, the two types of domains were seen to coexist, visualized by Raman mapping in the NCCDW phase (Fig.~1d). Thermal cycling to the ICCDW phase and back to the NCCDW phase left only $\beta$ domain (Fig.~1e), unchanged in subsequent cycles (Supplementary Fig.~2), because domain wall formation costs energy and is thermodynamically unstable. The planar chirality is preserved from bulk to atomically thin samples (Supplementary Fig.~3), suggesting that layer stacking forms a coherent domain state along the out-of-plane direction, allowing layer packing to accommodate the protruded or suppressed regions of the sulfur sub-layers~\cite{Fazekas1980,Spijkerman1997,Sung2022}.

\vspace{2mm}
\noindent
\textbf{Field-cooled poling}

The removal of coexisting domain states by thermal cycling indicates a small energy barrier between them, comparable to $k_{\mathrm{B}}T_{\mathrm{c}}$, where $k_{\mathrm{B}}$ is the Boltzmann constant and $T_{\mathrm{c}}$ the transition temperature between the NCCDW and ICCDW phases. This enables controlled switching between the two states, as discovered here, achievable by the remarkably simple application of a DC voltage. We found two switching pathways, as illustrated in Fig.~2a. 

We first demonstrate Pathway 1 for chirality switching. In analogy with ferroelectrics, we call this field-cooled poling, entailing cooling the sample from the high-symmetry to the low-symmetry phase under an external electric field. Figure~2b shows the optical image of a two-terminal device, which was initially in the $\alpha$ state (Fig.~2c). By maintaining a voltage while cooling from 360~K in the ICCDW phase (Supplementary Figs.~4 and 5), the state switched to $\beta$ when the voltage reached $-4.5$~V and alternated between $\alpha$ and $\beta$ for $V=\pm 4.5$~V. The $\beta$ state was verified to be metastable since the $\alpha$ state was initially favored. Figures~2d and 2e show the Raman mapping results at 4.5~V and $-4.5$~V, respectively. Each of the three regions separated by the electrodes was homogeneous. When Region 2 switched state at $-4.5$~V, Region 3 followed concomitantly, whereas Region 1 did not. Similar results were obtained in multiple devices (see another example in Supplementary Fig.~6), demonstrating controlled switching of the ferro-rotational order and domain-wall formation in close proximity to the electrodes. The tendency of the system to adopt a monodomain state implies strong interaction between contiguous regions, compromising the reproducibility of the switching. 
This can be mitigated by placing the electrodes on the perimeter of the flake or by separating unwanted regions.

\vspace{2mm}
\noindent
\textbf{Isothermal electrical switching}

Once a domain wall is introduced, isothermal switching of the ferro-rotational order can be achieved following Pathway 2. Figure~3a shows the device image for a 17.6~nm sample. Figure~3b shows Raman mapping of the target region, with the DC voltage applied in the shown order at 300~K and revoked for each measurement. The sample was in the $\beta$ state at 1~V. As the voltage decreased, the state switching initiated at the lower electrode and progressed upwards, until the entire region transformed to the $\alpha$ state at $-1$~V. The switching is therefore a result of domain wall motion. When the state switching completed, the domain wall was pushed to the immediate vicinity of one electrode, instead of being annihilated. Its shape changed as it swept across the channel, indicating that the seemingly sideways propagation entailed wall reconstruction.

Figure~3c shows the voltage dependence of $\Delta I_{\mathrm{int}}$ at a given location, measured at 270~K and 300~K. Abrupt switching above a critical bias $V_{\mathrm{c}}$ is revealed, and $|V_{\mathrm{c}}|$ is higher at lower temperature. This trend was confirmed at more temperatures (Supplementary Fig.~7), with $V_{\mathrm{c}}$ summarized in Fig.~3d, which reflects a decreasing domain wall mobility at lower temperature. Voltages exceeding large $|V_{\mathrm{c}}|$ induce significant Joule heating, which in turn lowers the threshold voltage, complicating the switching at low temperature (Supplementary Fig.~7). The temperature dependent $V_{\mathrm{c}}$ suggests that state switching can also be triggered by controlling temperature at fixed voltage. This is shown in Fig.~3e for $\pm 0.6$~V. The state only switches upon warming and does not reverse when cooled back. Indeed, the nonvolatility of the switched state is borne out in the electric hysteresis in Fig.~3c, showing retention of the switched state after voltage removal. The switching is also highly durable. Figure~3f demonstrates repeated cycles of direct voltage change between $\pm 1$~V at room temperature, successfully triggering state switching without any sign of fatigue (200 cycles tested). These results suggest applications in optoelectronics and nonvolatile memories.

Measurements using pulsed voltages offer insight into the mobility of the domain wall. Figure~3g shows that in a 10.0~nm thick sample forming an 8~$\mu$m channel, reproducible switching at 300~K was achieved for pulse duration longer than $\sim$10~ms, when the peak voltage was set to $\pm 1$~V (Supplementary Fig.~8). This amounts to a domain wall velocity of $10^{-4}$--$10^{-3}$~m/s, comparable to that in thin film ferroelectric Pb(Zr,Ti)O$_3$ measured under similar conditions~\cite{McGilly2015}. Figures~3h and~3i demonstrate reliable and deterministic switching using different patterns of voltage pulses.

Intriguingly, the state switching is accompanied by an abrupt small increase of the sample resistance (Fig.~4a). This is seen for both DC and pulsed voltage measurements (Supplementary Fig.~9). Figure~4b shows the temperature dependent resistance when the sample was cooled. The two monodomain states are degenerate, with close resistance values. The interim state with the domain wall exhibits a slightly higher resistance at 300~K, consistent with Fig.~4a, but a crossing occurs at about 220~K, yielding a lower resistance below that temperature. The Raman mapping results in the insets of Fig.~4b rule out cooling-induced change of the domain wall. The disparity between the monodomain and interim states is therefore ascribed to the domain wall. In the CCDW phase, imperfect SD clusters at the wall weaken the Mott insulating state~\cite{Ma2016,Cho2017,Skolimowski2019,Park2021}. The wall, expected to penetrate the entire sample thickness because of the interlayer coherence in each domain, may also affect the layer stacking order and break the interlayer dimerization~\cite{Ritschel2015,Ma2016,Lee2019,Butler2020}. Both factors lower the resistance. In the NCCDW phase, the increased resistance in the interim state indicates that the state switching not only affects the hexagonal patches containing closely packed SD clusters, but also extends into the conducting discommensuration regions~\cite{Park2019}, and the domain wall is more insulating than the discommensuration state.

\vspace{2mm}
\noindent
\textbf{Electrical switching mechanism}

The switching of the ferro-rotational order by a simple electric field is unexpected on symmetry grounds. The CCDW phase has well-defined superlattice periodicity with the $C_{3i}$ point group. The symmetry of the ferro-rotational order is $A_g$. Any conjugate field must be of the same symmetry to couple with this order~\cite{Cheong2018}, therefore excluding an in-plane electric field because of its $E_u$ symmetry. More complex forms of conjugate field, involving combinations of electric field, strain, and temperature gradient (see Supplementary Note 2), may qualify. The NCCDW phase lacks long-range order, lifting the symmetry requirement. Nonetheless, linear coupling with an electric field can still be ruled out: Flipping the sample about an axis containing the electric field changes the domain state as seen by the observer, but the electric field is invariant in this process. In other words, an electric field does not differentiate the two domain states. 

The electrical switching must then be ascribed to the action of the field on the domain wall, whose symmetry is much lower than a uniform domain. Specifically, we propose that the electric field acts upon the local charge dipoles at the wall to alter the local distortion, leading to the domain wall motion and state switching. Six S ions surround each Ta ion to form a symmetric TaS$_6$ octahedron in the undistorted phase. In the NCCDW/CCDW phase, the SD-type distortion causes off-centering of the Ta ions as well as displacement of S ions~\cite{Spijkerman1997}, inducing local charge dipoles. Figure~4c shows the local charge dipole in each octahedron for monolayer 1\textit{T}-TaS$_2$ obtained from first-principles calculations, which roughly aligns with the Ta displacement. Its magnitude is approximately 1~$e\cdot\mathrm{\AA}$ for the outer six octahedra in the SD cluster and smaller for the inner ones (Supplementary Fig.~10), similar to that in typical ferroelectrics~\cite{Catalan2012,Dawber2005}. There is no net global polarization because the local dipoles within a SD cluster cancel out, dictated by the centrosymmetric crystal structure. Figure~4d illustrates a possible form of wall between two different domains~\cite{Song2022}. An applied electric field can couple with the local dipoles and tends to orient them along the field direction, effectively nudging the ions such that they form a new SD and expand the domain along the field direction. This is consistent with the fact that two types of electric hysteresis loops were observed under the same configuration of the electrical circuit (Fig.~3c and Fig.~4a), as the electric field does not couple with a particular domain state, but only determines the propagation direction of the domain wall. The coercive field observed here is on the order of kV/cm, again similar to that in ferroelectrics~\cite{Dawber2005}, pointing to an analogous effect of the electric field on the domain wall in ferro-rotational 1\textit{T}-TaS$_2$ and typical ferroelectrics.

\vspace{2mm}
\noindent
\textbf{Discussion and conclusion}

Our results resemble the domain state switching in thick 1\textit{T}-TaS$_2$ flakes induced by femtosecond light pulses~\cite{Zong2018} in some aspects. Both types of switching can occur in the NCCDW phase but not in the CCDW phase. In that work, it was proposed that the NCCDW phase melts to yield a transient ICCDW phase before a different domain state emerges. Similarly, in our work we find that entering the ICCDW phase is essential for field-cooled poling of the domain states. According to electron diffraction, the ICCDW phase is characterized by disordered distributions of local $\alpha$- and $\beta$-type distortions~\cite{Sung2022}. Electric field can act upon the microscopic domain walls, as discussed earlier, assisting the growth of one type of domain when cooled into the NCCDW phase. On the other hand, the naturally formed NCCDW and CCDW phases are monodomain, which do not couple with electric field. The hysteretic electrical switching discovered here is distinct from the reversible switching of planar chiral states in monolayer 1\textit{T}-NbSe$_2$ nano-islands, achieved by voltage pulses of the same sign sent through the tip of a scanning tunneling microscope in the CCDW phase~\cite{Song2022}. The reason for the discrepancy in the switching conditions requires further studies. 

In summary, we have demonstrated hysteretic electrical switching of the ferro-rotational order in 1\textit{T}-TaS$_2$ by direct visualization of its domains and observation of domain wall propagation, establishing its ferroic character. The electric field does not couple with the order but only acts on the domain wall to trigger its motion. While the key microscopic mechanism for the switching has been outlined, a complete understanding of the process calls for further experimental and theoretical investigations. Whether  such a mechanism is universal is an open question. Our methodology is applicable to the discovery and investigation of more ferro-rotational materials.

\bibliographystyle{naturemag}
\bibliography{reference}

\vspace{3mm}
\noindent
\textbf{Methods}

\noindent
\textbf{Sample preparation}\\
Bulk TaS$_2$ single crystals were synthesized using the chemical vapor transport method. Thin flakes were mechanically exfoliated from the bulk crystals on silicone elastomer polydimethylsiloxane (PDMS) stamps in a glove box filled with high-purity argon gas, followed by transferring onto SiO$_2$/Si substrates and encapsulation with thin h-BN before being exposed to air. The flake thickness was determined by atomic force microscopy. 

\vspace{2mm}
\noindent
\textbf{Device fabrication}\\
For devices used in the electrical switching experiment and electrical transport measurements, thin TaS$_2$ samples on PDMS were transferred on pre-patterned Ti/Au electrodes and encapsulated by h-BN. The electrodes were fabricated by electron-beam lithography on Si substrates with a 300 nm SiO$_2$ layer, followed by electron-beam evaporation of 3~nm Ti and 15~nm Au. 

\vspace{2mm}
\noindent
\textbf{Laser cutting}\\
Sample cutting was realized by a supercontinuum laser coupled with an optical microscope, enabling simultaneous visualization of the sample and the laser spot. The laser emits pulses covering the visible to near-infrared spectral range with a pulse repetition rate of 0.2~MHz. The beam was focused on the sample by a 40$\times$ objective and the total incident power was greater than 10~mW. The sample was translated slowly using a manual translation stage, such that the trajectory under laser illumination was burnt due to excessive heat. A minimum linewidth of 1~$\mu$m was achieved. Successful cutting was indicated by change of sample color and confirmed by Raman measurements in the burnt region.

\vspace{2mm}
\noindent
\textbf{Raman measurements}\\
Raman micro-spectroscopy was performed in the back-scattering geometry using 532~nm laser excitation. The incident light was focused on the sample to a sub-micron-sized spot using an objective with high numerical aperture, 0.60 for regular measurements or 0.82 for Raman mapping. The scattered light was directed through Bragg notch filters before being collected by a grating spectrograph and a liquid-nitrogen-cooled charge-coupled device. The samples were mounted in a vacuum chamber during data acquisition. Raman mapping was performed by stepping the sample through a dense grid of coordinates using a piezoelectric two-dimensional scanner and collecting spectra at each position. 

The circularly polarized measurements were performed by using a combination of two polarizers and one quarter-wave plate, as detailed in Ref.~\cite{Yang2022}. Specifically, the first polarizer and the quarter-wave plate set the helicity ($\sigma^+$ or $\sigma^-$) of the incident light. The back-scattered light, containing both $\sigma^+$ and $\sigma^-$ components, passes through the quarter-wave plate again and is converted back to linear polarization. Setting the second polarizer either parallel with or perpendicular to the incident polarization therefore selectively probes the circularly contrarotating ($\sigma^+\sigma^-$ and $\sigma^-\sigma^+$) or the circulary corotating ($\sigma^+\sigma^+$ and $\sigma^-\sigma^-$) channels.

\vspace{2mm}
\noindent
\textbf{Temperature-dependent Raman and electrical transport measurements}\\
Temperature-dependent Raman and resistance data shown in Supplementary Fig.~3 were measured concomitantly in a Montana Instrument Cryostation. The temperature sweeping rate was set to a low value of 1.5~K/min because the CDW phase transitions in TaS$_2$ are sensitive to the cooling/warming rate~\cite{Yoshida2015}. Collection of both the Raman spectra (30~s/each) and the sample resistance were automated by a computer program. Sample drifting and defocusing of the laser spot were compensated by frequent manual adjustment. The resistance was measured in a four-probe geometry using the standard lock-in method, with an excitation current of 1~$\mu$A for the 103.7~nm sample and 0.1~$\mu$A for the 3.1~nm sample. The temperature-dependent two-probe resistance (Fig.~4b and Supplementary Fig.~9b) was measured using the same technique and cooling rate.

\vspace{2mm}
\noindent
\textbf{Electrical switching using DC and pulsed voltages}\\
Electrical switching of chirality was triggered by either DC or pulsed voltages, both in the two-terminal configuration. The devices were connected to the same external circuit to allow comparison of their characteristics. DC voltage was supplied by a Keithley 2400 Source Measure Unit, which also measured the induced current in the circuit, yielding the DC resistance $R_{\mathrm{DC}}$. Pulsed voltage was generated using a multifunction data acquisition device from National Instruments. Calibration of the pulse duration showed a minimum controllable value of $\sim$2~ms (Supplementary Fig.~8). For repeated switching using DC voltage shown in Fig.~3f, the voltage was held constant while Raman spectra were taken in both $\sigma^+\sigma^-$ and $\sigma^-\sigma^+$ channels (30~s/each); the completion of spectra acquisition triggered the change of voltage. Repeated switching using voltage pulses was carried out in a similar manner, with the voltage immediately revoked after one pulse duration and followed by Raman measurements. The pulsed-voltage dependence of the sample resistance shown in Supplementary Fig.~9a was measured using the lock-in method described above. Resistance was acquired after the passage of each voltage pulse. Relays were added to the circuits for electrical switching and for resistance measurement to avoid crosstalk between the two.

\vspace{2mm}
\noindent
\textbf{First-principles calculations}\\
The electric charge density before and after distortion of SD type is calculated by first-principles methods as implemented in Vienna \textit{ab-initio} Simulation Packages (VASP) \cite{Kresse1996}. Perdew-Burke-Ernzerhof (PBE) type generalized-gradient approximation (GGA) \cite{Perdew1996} is used for the exchange correlation functionals. A $5\times5\times1$ $k$-mesh is used for $\sqrt{13}\times\sqrt{13}$ supercell of 1$T$-TaS$_2$. The ground state energy and charge density data are collected as the total energy difference reaches the iterative threshold of $10^{-8}$ eV during the self-consistent calculation. The local charge dipole within the S octahedron surrounding each Ta ion is calculated as
\begin{equation}
    \mathbf{p} = 4e \mathbf{u}_{\textrm{Ta}} + \sum^6_{i=1} \left( -\frac{2e}{3} \right) \mathbf{u}^i_{\textrm{S}},\nonumber
\end{equation}
where $\mathbf{u}_{\textrm{Ta}}$ and $\mathbf{u}^i_{\textrm{S}}$ ($i=1,...,6$) represent the displacement vectors of the central Ta ion and the six surrounding S ions, respectively. Although the calculations were performed for the CCDW phase, the results are expected to represent the properties of the hexagonal patches in the NCCDW phase.

\vspace{3mm}
\noindent
\textbf{Data availability}\\
The data that support the plots in this paper and other findings of this study are available from the corresponding authors on reasonable request.

\vspace{3mm}
\noindent
\textbf{Acknowledgements}\\
We thank Junyao Yu and Chao Wang for assistance with the atomic force microscopy measurements. This work was supported by the National Key Research and Development Program of China (Grant Nos. 2018YFA0307000, 2017YFA0303201, and 2021YFA1400400) and the National Natural Science Foundation of China (Grant Nos. 11774151, 12225407, 12204160, and 12074174). K.W. and T.T. acknowledge support from JSPS KAKENHI (Grant Nos. 19H05790, 20H00354 and 21H05233) and A3 Foresight by JSPS. B.Y. acknowledges the financial support by the European Research Council (ERC Consolidator Grant ``NonlinearTopo'', No. 815869).

\vspace{3mm}
\noindent
\textbf{Author contributions}\\
X.X. conceived the project. G.L. and K.H. performed the measurements. T.Q. and D.L. fabricated the devices. Z.M., Z.H., and J.W. grew the TaS$_2$ crystals. K.W. and T.T. grew the h-BN crystals. G.L. and X.X. analysed the experimental data. Y.L. and B.Y. performed the DFT calculations. W.T, J.X., and L.G. performed atomic force microscopy measurements. X.X., B.Y., G.L., and J.M.L. interpreted the results. X.X. and B.Y. co-wrote the paper, with comments from all authors.

\vspace{3mm}
\noindent
\textbf{Competing interests}\\
The authors declare no competing interests.

\begin{figure*}[t]
\centering
\includegraphics[width=0.6\linewidth]{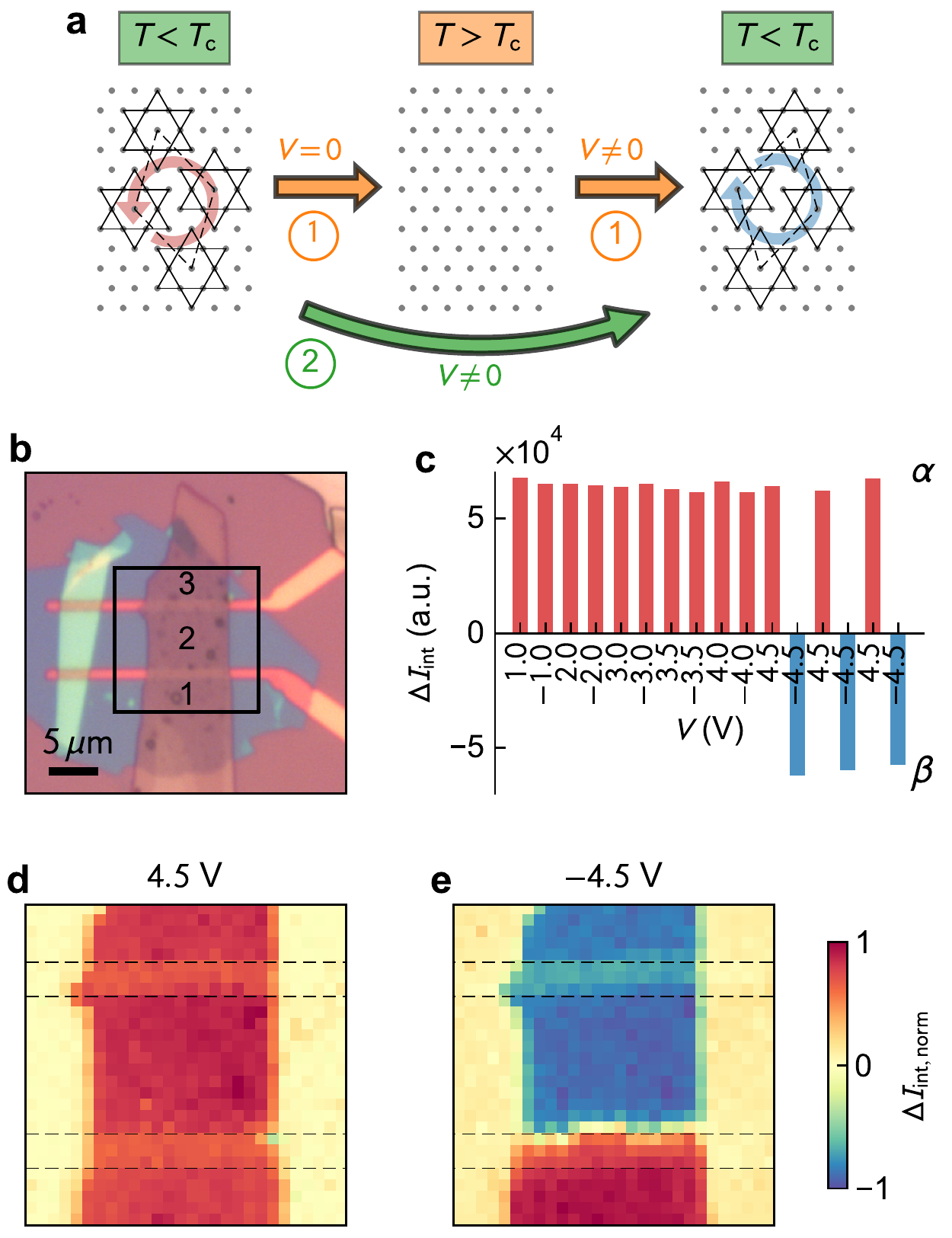}
\caption{\textbf{Electrical switching of ferro-rotational order via thermal cycling involving the ICCDW phase.} \textbf{a}, Illustration of two pathways for electrical switching. \textbf{b}, Optical image of a two-probe device for a 11.1~nm sample. Three regions are labeled. \textbf{c}, $\Delta I_{\mathrm{int}}$ taken at a fixed spot in Region 2 as the DC voltage is varied. \textbf{d}, \textbf{e}, Raman mapping of the region in the black square in \textbf{b}, after applying 4.5~V and $-4.5$~V, respectively. All data in \textbf{c}--\textbf{e} were taken at 300 K at zero voltage, following Pathway 1.}
\label{Fig2}
\end{figure*}

\begin{figure*}[t]
\centering
\includegraphics[width=1\linewidth]{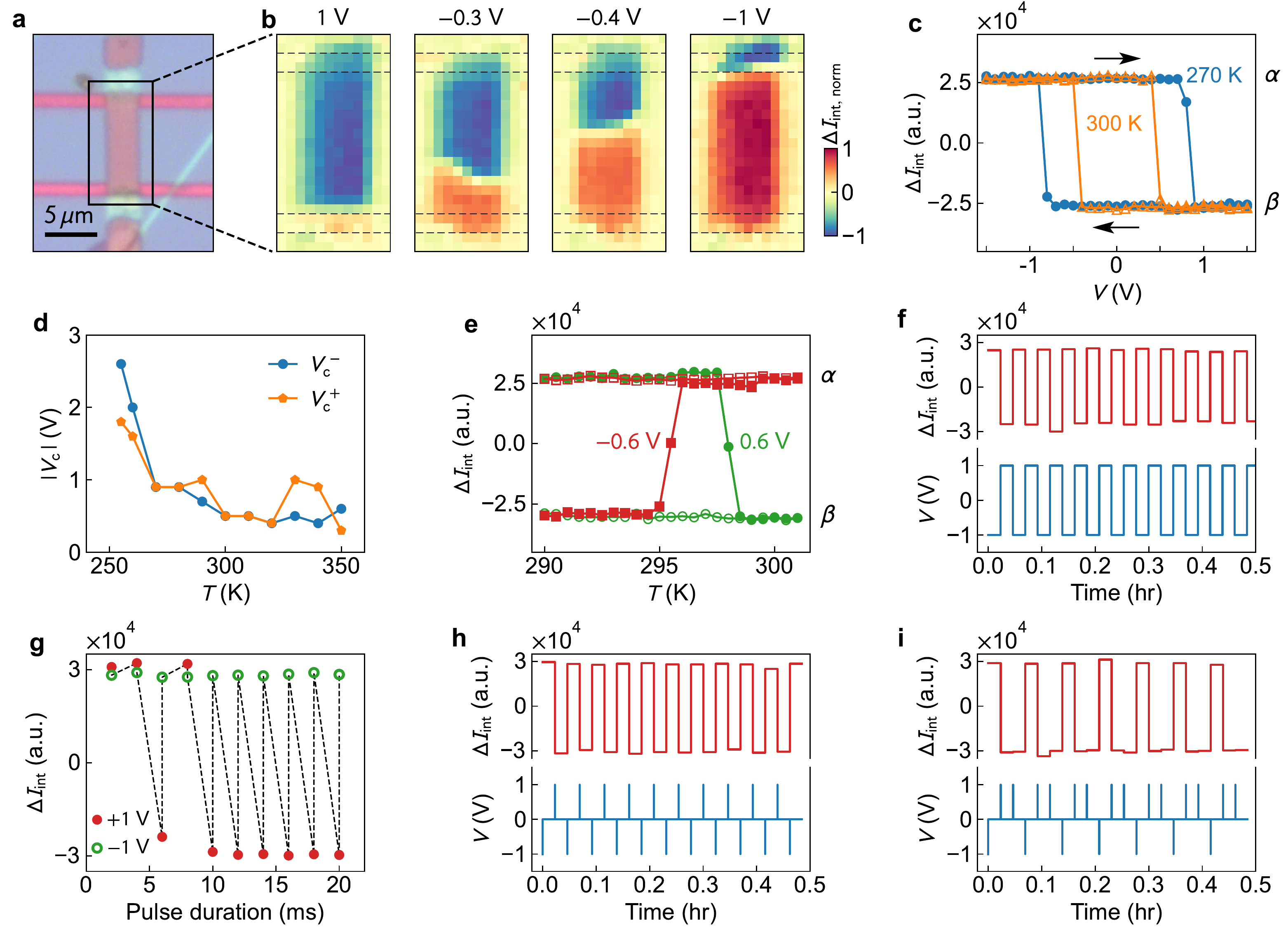}
\caption{\textbf{Isothermal electrical switching of ferro-rotational order in the NCCDW phase.} \textbf{a}, Optical image of a two-probe device for a 17.6~nm sample, with the parts outside the electrodes separated by laser cutting. \textbf{b}, Raman mapping of the marked region in \textbf{a} when the DC voltage was applied sequentially at 1, $-0.3$, $-0.4$, and $-1$~V at 300~K. \textbf{c}, DC voltage dependence of $\Delta I_{\mathrm{int}}$ at 270~K (circles) and 300~K (triangles). The arrows denote directions of voltage sweep. \textbf{d}, Temperature dependence of the critical voltage. \textbf{e}, Temperature dependence of $\Delta I_{\mathrm{int}}$ at $\pm 0.6$~V. Filled and open symbols represent warming and cooling, respectively. \textbf{f}, Repeated state switching using DC voltages.  \textbf{g}, Variation of $\Delta I_{\mathrm{int}}$ when voltage pulses with increasing pulse duration are applied, measured on a 10.0~nm device following the sequence indicated by the dashed lines. \textbf{h}, \textbf{i}, State switching using voltage pulses of 20~ms duration but different sequences of pulse trains. \textbf{c}--\textbf{i} were obtained with the laser spot at a fixed location. \textbf{f}--\textbf{i} are for 300~K.} 
\label{Fig3}
\end{figure*}

\begin{figure*}[t]
\centering
\includegraphics[width=0.9\linewidth]{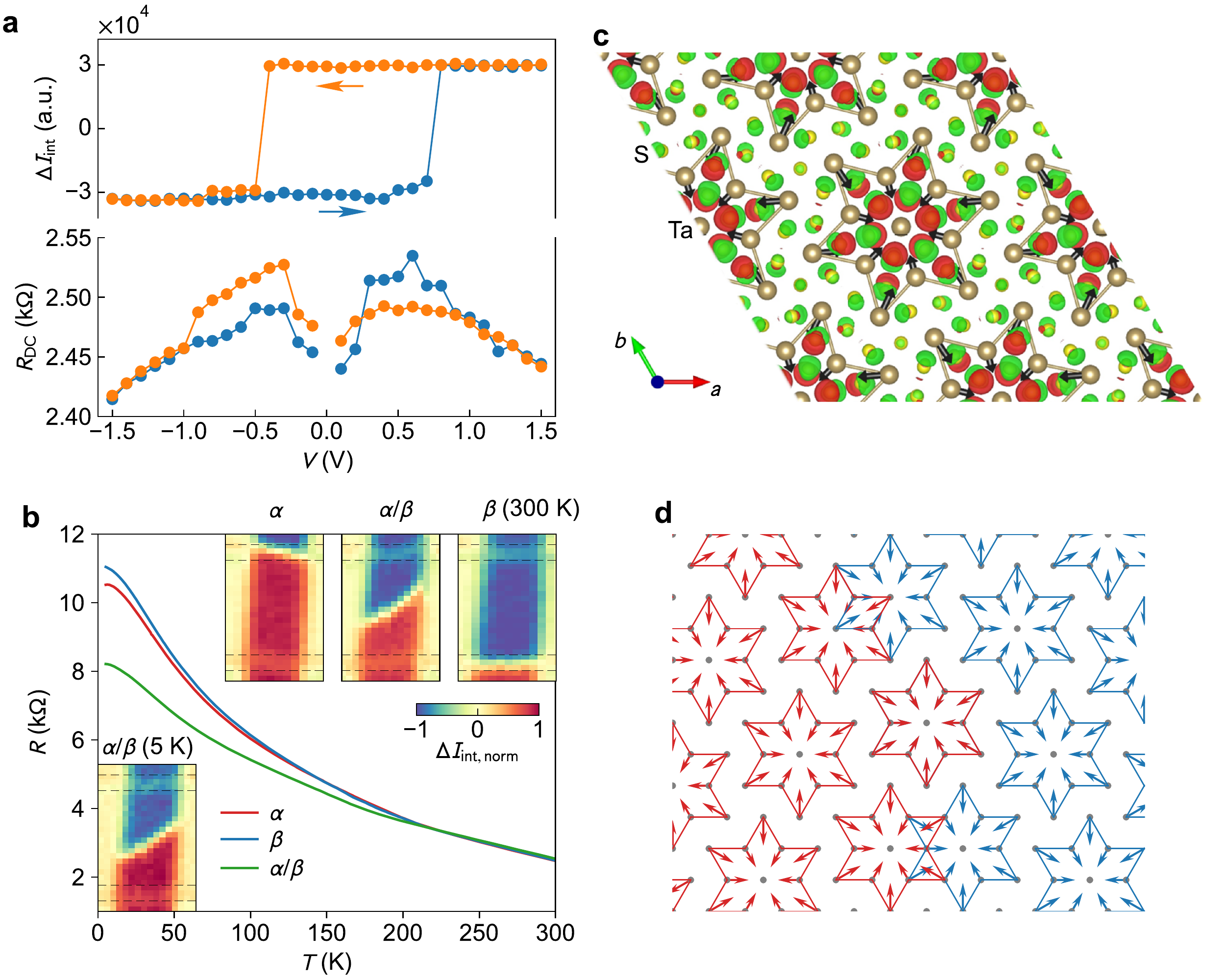}
\caption{\textbf{Electrical transport signature of the state switching and electrical switching mechanism.}  \textbf{a}, DC voltage dependence of $\Delta I_{\mathrm{int}}$ and DC resistance at 300~K. \textbf{b}, Resistance measured during cooling for different states. The insets are Raman mapping results. Dashed lines mark the electrodes, which are 8~$\mu$m apart. Sample thickness is 11.8~nm. \textbf{c}, Calculated charge redistribution due to the SD type CDW in monolayer 1\textit{T}-TaS$_2$. The red and green colors represent gain and loss of charges, respectively. The arrows indicate the local charge dipoles around each Ta atom. \textbf{d}, Schematic showing the formation of a domain wall between the two types of domains. Arrows represent the displacement of Ta ions in each SD cluster.} 
\label{Fig4}
\end{figure*}

\end{document}